# Improved accuracy of the NPL-CsF2 primary frequency standard: evaluation of distributed cavity phase and microwave lensing frequency shifts

Ruoxin Li[a], Kurt Gibble[a], and Krzysztof Szymaniec[b]

[a]*Department of Physics, The Pennsylvania State University, University Park, PA 16802 USA*
[b]*National Physical Laboratory, Hampton Road, Teddington, TW110LW, UK*

**Abstract**
We evaluate the distributed cavity phase and microwave lensing frequency shifts, which were the two largest sources of uncertainty for the NPL-CsF2 cesium fountain clock. We report measurements that confirm a detailed theoretical model of the microwave cavity fields and the frequency shifts of the clock that they produce. The model and measurements significantly reduce the distributed cavity phase uncertainty to $1.1 \times 10^{-16}$. We derive the microwave lensing frequency shift for a cylindrical cavity with circular apertures. An analytic result with reasonable approximations is given, in addition to a full calculation that indicates a shift of $6.2 \times 10^{-17}$. The measurements and theoretical models we report, along with improved evaluations of collisional and microwave leakage induced frequency shifts, reduce the frequency uncertainty of the NPL-CsF2 standard to $2.3 \times 10^{-16}$, nearly a factor of two lower than its most recent complete evaluation.

**1. Introduction**

The second, the unit of time in the International System of units SI, is defined in terms of a transition frequency between two ground state sublevels of an unperturbed atom of caesium 133. A precise determination of this frequency is, therefore, of fundamental importance in metrology and in timekeeping activities. National timescales, as well as the timescale disseminated by global navigation satellite systems, are normally steered in frequency to remain synchronized with the International Atomic Time (TAI) and with its derivative, the Coordinated Universal Time (UTC). The accuracy of TAI/UTC is assured by its step interval calibrations performed by primary frequency standards (PFS), which directly realize the SI second. Currently, the most accurate PFS are caesium fountains in which laser-cooled atoms are interrogated on their ascending and descending passages through a microwave cavity. The cavities have a high-Q resonance near the $^{133}$Cs ground-state hyperfine frequency. To date, several Cs fountain PFS have been constructed at a number of national timing laboratories and contribute to the steering of the TAI and UTC [1].

The accuracy of a particular PFS is estimated by evaluating uncertainties of all known systematic effects causing frequency shifts of the standard. Such effects are related to atomic interactions with external fields, collisions between atoms, and technical details of the construction of the standard's subsystems, such as the microwave cavity [2]. Over recent years, there has been a steady improvement in performance of the fountain PFS as more subtle systematic effects were minimized or evaluated with lower uncertainties. This has been possible thanks to simultaneous improvements in the reliability of operation, stability, and the development of an understanding of several physical effects.

At the National Physical Laboratory, atomic fountain PFS have been under development for several years [3],[4], with a first fountain standard contribution to TAI by NPL-CsF1 in 2005. A second caesium fountain, NPL-CsF2, became operational in 2006, and was upgraded and rebuilt in 2008 to exploit the cancellation of the collisional shift [5]. Compared with its predecessor, the environmental parameters of NPL-CsF2 are better controlled and its operation is more robust. Its first complete accuracy evaluation as a PFS gave a total type B uncertainty of $4.1 \times 10^{-16}$ [6]. The largest uncertainties were the frequency shifts from distributed cavity phase (DCP) and the mechanical forces that the microwave standing wave exerts on the atomic wavefunctions. This paper presents a quantitative evaluation of these previously leading uncertainties, and includes other recent improvements since [6].

**2. Distributed cavity phase frequency shift**

The DCP frequency shift is fundamentally a Doppler shift. The microwave field in the cavity is not a pure standing wave and, because the atoms move, both vertically and transversely, they see different phases of the microwave field during their upward and downward fountain traversals through the Ramsey cavity. A combination of analytic and finite-element models [7][8] has provided an apparently complete description of the cavity fields and the resulting DCP frequency shifts of the cylindrical TE$_{011}$ cavities that are used in nearly all primary fountain clocks [2]. A central feature of the model, for both calculating the fields and for evaluating the DCP uncertainties, is expanding the fields in a Fourier series in the azimuthal angle coordinate $\phi$. Each Fourier field component contributes to the spatial phase variation, produces a DCP frequency shift, and its symmetry dictates which physical effects contribute. Since the atoms are always near the cavity axis ($r\rightarrow0$), the fields and phase are proportional to $r^m\cos(m\phi)$ to lowest order, and therefore only a few terms, $m\leq 2$, contribute significantly [8]. Measurements at SYRTE have stringently confirmed the calculated fields and DCP shifts for their fountain [9]. Here we apply these models to the NPL-CsF2 fountain and present our measurements that further confirm the model and its description of the NPL-CsF2 DCP shifts.

The DCP frequency shift of any given PFS depends quite specifically on the cavity geometry and Q-factor, sizes and positions of the atomic cloud (e.g. MOT/molasses), and the particular apertures in the fountain. The Ramsey cavity in NPL-CsF2 is a cylindrical cavity with a nominal diameter and height of 43.0 mm, and a Q of 19,000. The endcaps have 10 mm diameter bores along the vertical axis. To precisely determine the cavity geometry, we measured a number of cavity resonances. For normal operation, the cavity's $TE_{011}$ resonance is tuned to the atomic clock's frequency within 30 kHz by controlling the temperature of the surrounding flight tube. The microwave power is supplied to the cavity through two rectangular waveguide transformers and coupled to the cavity by opposing 3 mm diameter holes in the cavity sidewalls at the midplane ($\phi = 0$ and $\pi$). Other inputs for the model of the fountain, for example, the cavity and aperture positions, were directly measured or calculated from the fountain geometry. We next describe the contributions from the three lowest Fourier terms.

The NPL-CsF2 Ramsey cavity is fed symmetrically for normal clock operation. This excites the Fourier field components with even $m$. The $m = 0$ component represents an azimuthally symmetric phase variation due to power flowing radially inward from the cavity midplane and being absorbed on the endcaps. This power flow implies very small transverse phase variations and large longitudinal phase deviations [7]. As a result, the DCP shift at low microwave amplitude $b$ is negligible (fig. 1a inset, solid curve) but the shifts are generally large at elevated amplitudes (fig. 1a, solid curve) [8]. Here $\delta P$ is the change in transition probability, $\delta P = -\delta \nu \, dP/d\nu$, where $\delta \nu$ is the frequency shift and $dP/d\nu$ is the Ramsey fringe slope, approximately $-\pi/2\Delta\nu^{FWHM}$ at optimal amplitude and a detuning of $\Delta\nu^{FWHM}/2$, where $\Delta\nu^{FWHM}$ is a full linewidth of the fringe. The microwave amplitude at $b=1$ corresponds to a $\pi/2$ pulse on average for atoms uniformly illuminating the cavity aperture. Because the atomic cloud is small and centred on the upward passage, the atoms on average experience more than a $\pi/2$ pulse at $b=1$, and thus $(n=1,3,5…)\times\pi/2$ pulses (black dots), where the Ramsey fringe contrast is largest, occur at $b$'s somewhat less than 1,3,5… We use these points and the measured Ramsey fringe contrast to determine the microwave amplitude $b$. In the inset in fig. 1a, the calculated $m=0$ DCP frequency shift is $1.6 \times 10^{-17}$ at $n=1$ and we correct the clock for this bias. The measurements in fig. 1c do not exclude that the endcaps could have different surface conductivities, at the level of 20% [9]. Such a difference would produce an $m=0$ DCP

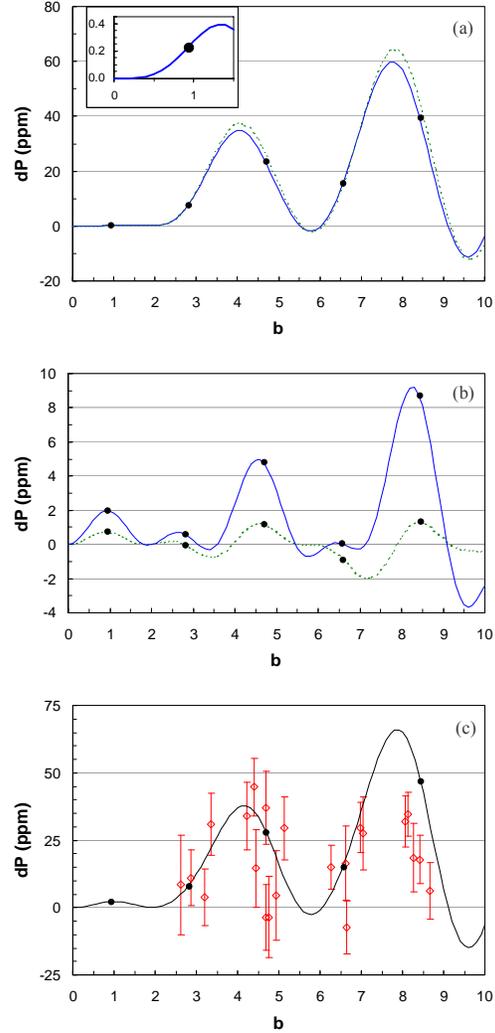

Fig. 1. DCP frequency shifts of even azimuthal modes: (a) Predicted $m=0$ contribution to the change in transition probability $\delta P$ as a function of the scaled microwave field amplitude $b$, described in the text. The solid (dashed) curve is for an atomic cloud launched on (1.1 mm off) axis with a 7 mm beam waist detection laser propagating perpendicular to the feeds. The black dots correspond to the locally maximum Ramsey fringe contrast for $(1,3,5,…) \times \pi/2$ pulses. The inset shows $\delta P$ near optimal field amplitude ($b \approx 1$); the two curves are indistinguishable. (b) Predicted $m=2$ DCP shifts. The curves and dots are as in (a), but with uniform state detection for the 1.1 mm offset cloud (dashed curve). (c) Measured (open diamonds) and predicted (solid curve) DCP shifts. The predicted curve is the sum of the $m=0$ and $m=2$ solid curves in (a) and (b).

shift of $1.6 \times 10^{-17}$ at $n=1$, which we take as the $m=0$ DCP uncertainty.

Quadrupolar phase variations ($m=2$) may also produce a significant DCP frequency bias for cavities with only two feeds [8][9]. Both cloud offsets and non-uniform state detection produce $m=2$ DCP shifts. Our state detection laser beams propagate perpendicularly to the feeds and have a Gaussian transverse intensity profile with a $1/e^2$ radius of 7 mm. Assuming a cloud launched on the cavity axis (fig. 1b, solid curve) and spatially uniform fluorescence collection, the model gives a frequency shift of $13.8 \times 10^{-17}$. We correct the fountain frequency for this bias and take half of its value as the uncertainty. We estimate that the fluorescence collection has a Gaussian $1/e^2$ radius of at least 2 cm, which would reduce the calculated shift by less than 10%. An additional and uncorrelated uncertainty of $3.6 \times 10^{-17}$ (fig. 1b, dashed curve) follows from a 1.1 mm cloud position uncertainty (see below). Combined, these give a total $m=2$ DCP uncertainty of $7.8 \times 10^{-17}$ (table 1).

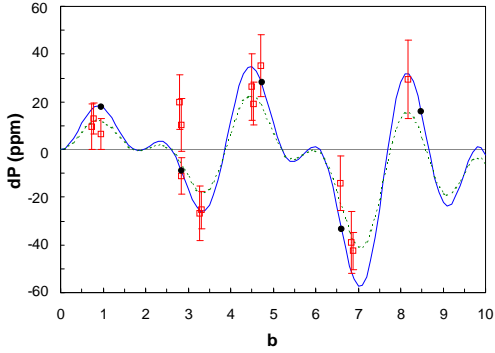
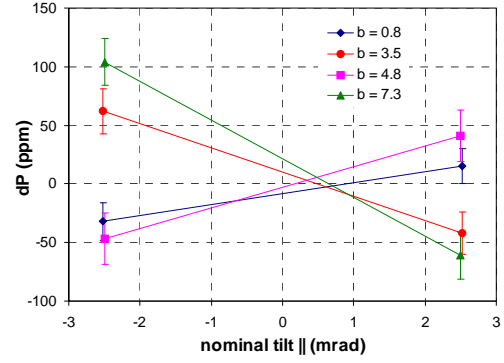

Fig. 2. Predicted and measured $m=1$ DCP shifts. We take half the frequency difference between feeding the cavity alternately at $\phi=0$ or $\pi$ with a 2.5 mrad fountain tilt (solid curve and open squares). The dashed curve is for no tilt and a 1.1 mm offset of the launched cloud from the fountain axis. Black dots are as in fig.1.

Fig. 3. The $m=1$ DCP shift measured by feeding alternately at $\phi=0$ or $\pi$ as in fig. 2, for several field amplitudes $b$, versus fountain tilt. The tilts were nominally $\pm 2.5$ mrad. Linear fits determine the actual vertical to be $(-0.55 \pm 0.33)$ mrad.

Table 1. DCP frequency shifts of azimuthal Fourier terms.

| azimuthal term | shift / $10^{-16}$ | uncertainty / $10^{-16}$ |
|---|---|---|
| $m = 0$ (calculated) | 0.16 | 0.16 |
| $m = 1$ ∥ (measured) | - | 0.61 |
| ⊥ (measured) | - | 0.47 |
| $m = 2$ (calculated) | 1.38 | 0.78 |
| total DCP frequency shift | 1.54 | 1.11 |

In fig. 1c the fountain's frequency was measured with respect to optimal amplitude ($n=1$) for $3 \leq b \leq 9$. The fountain was operated alternately at elevated and optimal amplitudes and the data are corrected for the small $m=2$ shift at optimal power ($n=1$). Significant changes in the fountain's frequency at elevated amplitudes could result from other effects, including microwave leakage [10], spurious components in the microwave spectrum [11], and collisional shifts [12]. We have previously shown that neither the leakage nor the spurs produce measurable frequency shifts of NPL-CsF2 [6], but the collisional frequency shift does depend on $b$. We therefore measure and correct for the collisional shift at each $b$, additionally alternating between high and low atomic densities. This significantly increases the statistical error bars, making them comparable to the predicted shifts. The data and model agree, with some systematic deviation for $b>8$, where fast longitudinal spatial phase variations contribute. We cannot exclude that these could be caused by an inhomogeneous cavity wall resistance [9].

The $m=1$ field component represents a phase gradient and power flow from one side of the cavity to the other, e.g. $\phi=0$ to $\phi=\pi$. Feed imbalances create an $m=1$ field component and it essentially only produces a DCP shift if the centres of mass for the ascending and descending atomic clouds are displaced. Tilting the fountain away from vertical, by several mrad, and feeding the cavity with only one feed, alternately $\phi=0$ or $\pi$, gives measureable frequency differences in fig. 2 [9][13]. Again, the model predicts a significant dependence on microwave amplitude $b$, which we observe. At optimal amplitude the frequency shift is $1.25 \times 10^{-15}$ ($\delta P = 18$ ppm) for a 100% imbalance in the feed amplitudes and a 2.5 mrad tilt of the fountain. For normal clock operation, the m=1 DCP shift is minimized by balancing the feeds and minimizing the tilt.

The measurements of $m=1$ DCP frequency shifts in fig. 2 are due to the feeds and do not include possible phase gradients from inhomogeneous wall losses or differences in the reflectivities of the feeds [9]. To account for these, we consider two cases for the $m=1$ phase gradients: parallel and perpendicular to the feed axis, and we measure the tilt sensitivity of the fountain frequency $\Delta v^{tilt}$ in these two orthogonal directions as in [9]. We initially balance the feeds by adjusting them to give the same pulse areas ($\pi/4$) on the upward passage. We then check the tilt sensitivity at optimal power using the NPL hydrogen maser HM2 as a frequency reference.

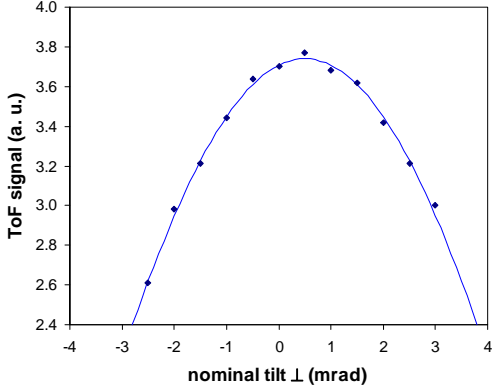

Fig. 4. Integrated detection time-of-flight signal (diamonds) as a function of the nominal tilt of the fountain, perpendicular to the microwave cavity feed axis. To precisely determine the tilt angle that maximizes the detection signal, a parabola was fit (solid curve).

Its instability remained below the CsF2 uncertainty for averaging times of more than one day, and therefore, by alternating the tilt between two extreme values, ±3.0 mrad, for many days, the maser long-term drift is rejected. Within the statistical uncertainty, no significant tilt sensitivity was found in either tilt direction: $\Delta v^{tilt}_{\parallel} = (1.2 \pm 1.4) \times 10^{-16}$ mrad$^{-1}$ and $\Delta v^{tilt}_{\perp} = (1.1 \pm 1.1) \times 10^{-16}$ mrad$^{-1}$. If there was a significantly nonzero value of $\Delta v^{tilt}_{\parallel}$, it would be minimized by balancing the feed amplitudes [9].

We null the tilt of the fountain for phase gradients along the feeds using differential measurements as in fig. 2, for several field amplitudes at two large tilts (fig. 3) [9][13]. The zero-crossing in fig. 3 defines vertical for $m=1$ DCP shifts with an uncertainty of ±0.33 mrad. We note that the tilt offset in fig. 3 is much larger than the 10 µrad resolution of our precision spirit levels. The tilt uncertainty, multiplied by the quadrature sum of $\Delta v^{tilt}_{\parallel}$ and its uncertainty, gives our $m=1$ DCP uncertainty for phase gradients along the feeds of ± 6.1×10$^{-17}$.

For tilts perpendicular to the feeds, we cannot null the tilt using measured $m=1$ DCP shifts as above with our current two-feed Ramsey cavity. Instead, we determine the initial position of the MOT by carefully aligning the vertical laser cooling beams using small diaphragms and the retroreflection from the surface of a liquid. We find that the cloud is launched on the vertical axis of the fountain to within 1 mm. After subsequent realignment or rebalancing of the cooling beams, the atomic cloud's position changed by less than 0.5 mm, as observed with a triggered camera [14]. Adding these two position uncertainties in quadrature gives an offset uncertainty of 1.1 mm. Since the descending atomic cloud is centred in the detection region because the number of detected atoms is maximized by adjusting the fountain's tilt (fig. 4), we obtain a perpendicular tilt uncertainty of ±0.3 mrad. This value is coincidentally similar to the uncertainty of the parallel tilt obtained from the interpolations of the tilt sensitivity (fig. 3). Adding the parallel and perpendicular $m=1$ DCP uncertainties in quadrature gives our total $m=1$ DCP uncertainty of ±7.7 × 10$^{-17}$. Table 1 summarizes the significant DCP error contributions for NPL-CsF2 that give an overall DCP frequency shift uncertainty of ±1.1 × 10$^{-16}$.

The phases of the two cavity feeds are balanced to within 10 mrad using the atoms as a probe. When the cavity is tuned near resonance, the DCP frequency shift from phase imbalances is suppressed [8][9]. The NPL-CsF2 cavity tuning is maintained within $\Gamma/15$ of resonance, where $\Gamma$ is the resonance full width. The suppression of the DCP shifts is 1/1500 of that for a single feed, giving a negligible DCP shift of $1 \times 10^{-18}$. Note that this negligible shift is included in the measurement of $\Delta v^{tilt}_{\parallel}$ above.

A DCP shift could be produced by potential machining burrs in the cutoff waveguides that could produce locally large fields and modify the density distribution [8][9]. However, the small initial atomic cloud size in NPL-CsF2 and the lower cavity aperture limit atoms from travelling within 100 µm of the "corners" formed by the cutoff waveguides and the cavity endcaps. Here we consider atoms launched from the 1/$e$ width of the cloud when the cloud is displaced from the cavity axis by 1.1mm, and a fountain tilt of 0.33 mrad. Atoms that are closer than 100 µm to the corners do not pass through the lower selection cavity aperture [6]. Since no burrs are believed to extend 100 µm into the cutoff waveguides, potential burrs could not produce a DCP shift [8].

**3. Microwave lensing frequency shift**

The microwave standing-wave in the Ramsey cavity exerts mechanical dipole-forces on the atomic wavefunctions as they pass through the cavity. The field acts as a weak positive (negative) lens on the wavefunctions of the atomic dressed states $|2(1)\rangle$ during the first cavity traversal [15]. At the second Ramsey cavity passage, the field is phase shifted by $-(+)\pi/2$ and transfers dressed state $|1(2)\rangle$ to the final state of the clock transition. Thus, for a positive detuning, more final-state atomic population passes through the cavity aperture than for a negative detuning, producing a positive frequency shift of the Ramsey fringe. Another treatment described this shift as the recoil shift from a microwave photon [16]. However, the atomic wavefunctions are restricted to much less than a microwave wavelength. Thus, there are not discreet recoils but instead a lensing of the wavefunctions, which in turn implies that apertures in the fountain are essential for a proper treatment of this frequency shift. Nonetheless, the size scale for this frequency shift is of order of the recoil shift for a microwave photon [15]. This simplistic, usually conservative, estimate was used for the first accuracy evaluation of the NPL-CsF2 and, with the improved DCP uncertainty above, it would be the largest contribution to its uncertainty. Here we explicitly extend the results of [15] to describe circular cavity apertures, and also include the higher order terms and detection non-uniformities.

It is helpful to first give an approximate analytic result, especially since it is often reasonably accurate. Ref [15] gives a simple analytic result for rectangular apertures and cylindrical or rectangular cavities, for dipole forces in just one or both transverse directions. The approximations for that analytic result were: terms only to second order in the microwave wave vector $k$, uniform detection, and neglecting the clipping by the cavity apertures on the upward passage and the decrease of the Ramsey fringe contrast at high amplitudes. The physical behaviours in [15] are unchanged for circular apertures and a cylindrical cavity, for which we get, with the same approximations:

$$\frac{\delta v}{v_R} = \frac{b_1 \eta \pi}{2\sin\left(b_1 \eta \frac{\pi}{2}\right)} \frac{a^2\left(w_0^2 + t_1 t_{2L} u^2\right)}{w_{2L}^4 \left(e^{\frac{a^2}{w_{2L}^2}} - 1\right)} \frac{(t_{2L} - t_1)}{(t_2 - t_1)}. \quad (1)$$

Here the usual recoil frequency shift is $v_R = hv^2/2m_{Cs}c^2$, $a$ is the radius of the cavity apertures, $w_0$ is the $1/e$ cloud radius at launch, $u=(2k_B T/m_{Cs})^{1/2}$ is the thermal velocity, $t_{1,2,2L}$ are the times at which the atoms pass through the Ramsey cavity and the lower cavity aperture on the downward passage, $w_{2L}^2 = w_0^2 + u^2 t_{2L}^2$ gives the final cloud size at the lower cavity aperture, $b_{1,2}$ is the scaled microwave amplitude, and $\eta=1.120$ for $a=5$mm [8]. Eq. (1) gives $\delta v/v=6.25\times10^{-17}$ for NPL-CsF2, about 40% of the simplistic microwave recoil shift, with $a=5$mm, $w_0=1.1$mm, $u=15$mm/s, $t_1=0.18$s, $t_2=0.7$s, $t_{2L}=0.837$s, and $b_1 = 0.9386$.

Our treatment begins with the change in transition probability [15][17],

$$\delta P = \frac{1}{2N}\int\left[\left|\langle 2|\Psi\rangle\right|^2 - \left|\langle 1|\Psi\rangle\right|^2\right]\sin\left[\theta(r_2)\right]W_d(\vec{r}_d)dz_2 d\vec{r}_2$$

including the detection probability $W_d(\mathbf{r}_d)$. Here $N$ is the number of detected atoms defined below, all vectors are transverse, $\theta(r_2) = (\pi/2)b_2\eta J_0(k\ r_2)$ is the position-dependent tipping angle of the downward Ramsey pulse at transverse position $r_2$ [8], and $k=2\pi v/c$. We can consider the wavefunction for each atom as being localized to within an optical wavelength when launched by the molasses cooling light [15]. Thus, the wavefunctions are from many uncorrelated, effectively point sources, each with a velocity spread, $W_T(v)$, given by the temperature. We can therefore use a semi-classical propagation of the wavefunctions that are straight-line trajectories, except for a deflection during the first Ramsey interaction. We integrate over the trivial $z_2$ dimension and the initial transverse source's spatial distribution $W_{r_0}(\mathbf{r}_0)$. We then change variables so that the integrations are over the transverse velocity and the aperture at the first Ramsey interaction $\mathbf{r}_1$.

$$\delta P = -\frac{(t_{2L}-t_1)^2}{4N}$$
$$\times \sum_{\pm} \pm \int\int_{r_1<a,\ r_{1L}<a} W_T(\vec{v})W_{r_0}(\vec{r}_0)\sin\left[\theta(r_2)\right]W_d(\vec{r}_d)d\vec{v}d\vec{r}_1,$$
$$\vec{r}_\beta = \vec{r}_1 + \left[\vec{v}\pm\delta\vec{v}(\vec{r}_1)\right](t_\beta - t_1) \quad \beta\in\{2, 2L, d\}, \quad (2)$$
$$\vec{r}_0 = \vec{r}_1 - \vec{v}t_1,\quad \vec{r}_{1L} = \vec{r}_1 - \vec{v}(t_1 - t_{1L})$$

Here $+(-)$ corresponds to dressed states $|1(2)\rangle$, $t_{1L}$ is the time the atoms pass the lower cutoff waveguide aperture on the upward traversal, and $\delta v(\mathbf{r}_1)$ is the position-dependent velocity change during the upward cavity traversal, which includes terms to all orders in $k$,

$$\delta\vec{v}(\vec{r}_1) = -b_1\eta\pi^2 \frac{v_R}{k^2}\vec{\nabla}J_0(kr_1). \quad (3)$$

It is convenient to now change variables from velocity to the undeflected position of an atom at the lower aperture on the downward passage, $\mathbf{r}_{2L0} = \mathbf{r}_1 + \mathbf{v}(t_{2L}-t_1)$.

We expand (2) and keep terms only to first order in $v_R$. We get a line-integral around the aperture at $t_{2L}$ and a surface integral over that aperture, with both being integrated over the surface $\mathbf{r}_1$.

$$\delta P = \frac{a}{2N}(t_{2L}-t_1)\int_{r_1<a}\left|\delta\vec{v}(\vec{r}_1)\right|\int_0^{2\pi}P(\vec{r}_1,\vec{r}_{2L0})$$
$$\times\sin\left[\theta(r_2)\right]\Big|_{\substack{r_{2L0}=a \\ \delta v(\vec{r}_1)=0}}\cos(\phi_{2L0})d\phi_{2L0}d\vec{r}_1 \quad (4)$$
$$+ \frac{v_R}{2N}\int_{r_1<a}\int_{r_{2L0}<a}\frac{\partial P(\vec{r}_1,\vec{r}_{2L0})\sin\left[\theta(r_2)\right]}{\partial v_R}\bigg|_{\delta v(\vec{r}_1)=0}d\vec{r}_{2L0}d\vec{r}_1$$
$$P(\vec{r}_1,\vec{r}_{2L0}) = W_T(\vec{v})W_{r_0}(\vec{r}_0)W_d(\vec{r}_d)\Theta(a-r_{1L})$$
$$N = \int\int_{r_1<a,\ r_{2L0}<a}P(\vec{r}_1,\vec{r}_{2L0})\Big|_{\delta v(\vec{r}_1)=0}d\vec{r}_{2L0}d\vec{r}_1$$

Here $\cos(\phi_{2L0}) = \hat{r}_1 \cdot \hat{r}_{2L0}$, and $\Theta(x)$ is the Heaviside step function that describes the lowest aperture in the fountain, and any others before $t_1$. When the fountain is tilted, apertures after $t_1$ may clip the atomic cloud, modifying both terms in (4). We note that $W_d(\mathbf{r}_d)$ and $\sin[\theta(r_2)]$ depend explicitly on $v_R$ via $\delta v(\mathbf{r}_1)$. In this way, Eq. (4) gives the full microwave amplitude and tilt dependent microwave lensing. The Ramsey fringe amplitude $\Delta P_R$ can be calculated from integrating $P(\mathbf{r}_1,\mathbf{r}_{2L0})\sin[\theta(\mathbf{r}_1)]\sin[\theta(r_2)]$, as for $N$ in (4), to give the microwave lensing frequency shift $\delta v = \delta P/\pi(t_2-t_1)\Delta P_R$.

We now evaluate (4) and successively introduce corrections. First, we simply consider optimal amplitude, where $\sin[\theta(r_2)]$ is very flat. With $W_d(\mathbf{r}_d)=1$, the second term in (4) gives no contribution since the derivative is 0. We can take Gaussian velocity and initial position distributions, $W_T(v)=\exp(-v^2/u^2)$ and $W_{r_0}(\mathbf{r}_0)=\exp(-r_0^2/w_0^2)$, and consider $w_0 \ll a$ so that the integral over $r_1$ is over all space. Taking only terms of order $k^2$ in $\delta v(r_1)$ gives (1). Restricting the integral over $r_1$ to $r_1<a$ increases the frequency shift only slightly, to $6.3\times10^{-17}$ since $w_0$ is much less than $a$.

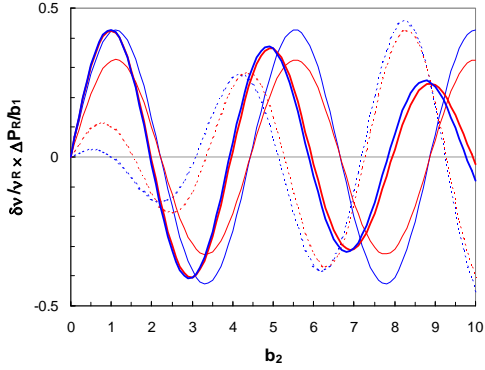

Fig. 5. Microwave lensing frequency shift versus microwave amplitude $b_2$ of the second Ramsey pulse. Curves are shown for uniform detection (blue) and a Gaussian laser beam with 7mm waist (red). The total shifts are the thick lines, and the first (second) terms in (4) are thin (dashed). The shift, $\delta\nu\,\Delta P_R$, or equivalently $\delta P$, increases linearly with the amplitude $b_1$ of the first microwave Ramsey pulse.

Terms higher order than $k^2$ contribute to both the velocity change in (3) and to $\sin[\theta(r_2)]$ of $\delta P$ in (4). The small cloud size $w_0$ of NPL-CsF2 means this correction is small for $\delta v(r_1)$ and it is always small for $\sin[\theta(r_2)]$ at optimal power. The first term of (4) gives $6.02\times10^{-17}$, the total is $6.14\times10^{-17}$, and these are unchanged when the lower aperture $r_{1L}<a$ is included. At elevated microwave amplitude, the second term in (4) is important, especially near $b_2\approx4.5$ where $\sin[\theta(r_2\approx a)]$ and the first term in (4) goes to zero (fig 5, dashed line). Including the detection variation from the Gaussian detection beam (7 mm radius waist) gives a slightly smaller effective aperture. The total lensing frequency shift is nearly the same, $6.20\times10^{-17}$, but now the second term in (4) has a larger contribution, $1.59\times10^{-17}$. In fig. 5, the detection non-uniformity has little effect on the overall shift, but it significantly changes the two contributions from (4). We correct the clock's frequency for the microwave lensing frequency shift. Even though the various corrections are very small, we nonetheless take half of its value, $3.1\times10^{-17}$ in table 2, as its uncertainty, in part, because it has not yet been experimentally observed.

## 4. Other improvements

The NPL-CsF2 fountain standard operates in the vicinity of zero-collisional shift [18]. To correct for a possible residual shift, the fountain is run alternately at high and low atom number and the measured frequency is extrapolated to that for zero atomic density. We distinguish between type A and type B errors of the collisional shift. The type A error is included in the total type A uncertainty of the standard and decreases as $\tau^{-1/2}$, where $\tau$ is the averaging time. The type B error originates from a systematic uncertainty of the ratio $\kappa$ of the high to low density, as only the detected atom number ratio is measured in the fountain, with $\kappa$ known to 10%. The recent implementation of an optical pumping stage in NPL-CsF2, which accumulates population of one of the clock states, enabled the operation of the standard at higher $\kappa$. As a result, the collisional shift uncertainty is less sensitive to the uncertainty of $\kappa$ [14]. For example, for $\kappa\approx8$ and a measured frequency difference between high and low density of $<2.5\times10^{-15}$, the type B uncertainty of the collisional shift is less than $4\times10^{-17}$.

In an earlier evaluation, the uncertainty from microwave leakage was dominated by the potential leakage, through the switches that pulse the field for the upward passage, into the selection cavity, which could possibly perturb the atoms when they descend. A recently repeated evaluation of this effect gave a smaller statistical error. In addition, after implementing the optical pumping and optimising the fountain operation [14], the fountain runs most of the time at low atomic density, and hence at a low microwave amplitude in the selection cavity. As a consequence, our uncertainty is smaller, of $3\times$ $10^{-17}$, for this frequency shift. Together with the unchanged uncertainty of a frequency shift due to the leakage of the interrogating field to the area outside the cavities ($5 \times 10^{-17}$), the total leakage frequency shift is now less than $6 \times 10^{-17}$.

Table 2. Uncertainty budget for NPL-CsF2. The underlined systematic effects have reduced uncertainties as compared to [6].

| Type B evaluation | uncertainty / $10^{-16}$ |
|---|---|
| *Effect* | |
| Second order Zeeman | 0.8 |
| Blackbody radiation | 1.1 |
| AC Stark (lasers) | 0.1 |
| Microwave spectrum | 0.1 |
| Gravity | 0.5 |
| Cold collisions (Cs-Cs) | 0.4[a] |
| Collisions with background gas | 1.0 |
| Rabi, Ramsey pulling | 0.1 |
| Cavity phase (distributed) | 1.1 |
| Cavity phase (dynamic) | 0.1 |
| Cavity pulling | 0.2 |
| Microwave leakage | 0.6 |
| Microwave lensing | 0.3 |
| Second-order Doppler | 0.1 |
| $u_B$ (1$\sigma$) | 2.3 |
| Type A evaluation | |
| $u_A$ (1$\sigma$, for 15-day averaging) | 2.4 |
| Total uncertainty | 3.3 |

[a] An exemplary value of the type B contribution to the uncertainty for a measured residual collisional shift, the frequency difference between high and low density, below $2.5\times10^{-15}$ [14].

## 5. Summary

We have performed detailed evaluations of the distributed cavity phase and microwave lensing frequency shifts. These were the two largest systematic uncertainties in our most recent evaluation of the NPL-CsF2 primary frequency standard. Our measurements of DCP frequency shifts add further support to the validation in [9] of the recent theoretical model [8]. The model, with no free parameters, gives predicted m=0 and 2 DCP shifts, which we use to bound their uncertainties. Uncertainties for m=1 DCP shifts are experimentally determined from measurements of the fountain's frequency versus the tilt of the entire apparatus to probe phase gradients along and perpendicular to the cavity feed axis [9]. These shifts could arise from feed assymetries or a non-uniform surface resistance of the cavity's copper walls. The residual tilt uncertainty and the non-uniformity of the detection beam provide the largest contributions to the total DCP uncertainty of $1.1 \times 10^{-16}$. We calculate the frequency shift due to the mechanical action of the microwave standing wave on the atomic spatial wavefunctions by extending the theoretical model of [15]. The predicted microwave lensing frequency shift is $6.2 \times 10^{-17}$, 40% of a simplistic microwave recoil shift. The importance of this shift in the uncertainty budgets demonstrates that the best Cs fountain clocks are, in a clear physical interpretation, matter-wave interferometers. Finally an optical pumping stage increased our clock state populations and enabled a further reduction of the type B uncertainty of the residual collisional shift. This analysis and these improvements to the NPL-CsF2 clock lead to a nearly two-fold reduction of its type B uncertainty, to $2.3 \times 10^{-16}$.

**Acknowledgements**

Contributions of Sang Eon Park and Akifumi Takamizawa to the NPL experimental work are gratefully acknowledged. We acknowledge financial support from the NSF, Penn State, and the UK National Measurement Office.